\documentclass[sigconf,nonacm]{acmart}
\AtBeginDocument{%
  }
\setcopyright{acmlicensed}
\copyrightyear{2026}
\acmYear{2026}
\acmDOI{10.1145/3773078.3831800}
\acmConference[RecSys '26]{20th ACM Conference on Recommender Systems}{September 28--October 2, 2026}{Minneapolis, MN, USA}
\acmISBN{979-8-4007-2284-4/26/09}

\usepackage{booktabs}
\usepackage{multirow}
\usepackage{xspace}
\usepackage{graphicx}
\usepackage{algorithm}
\usepackage{algorithmic}
\usepackage{relsize}
\usepackage{xcolor}
\newcommand{\ci}[2]{{\scriptsize\textcolor{gray}{[#1, #2]}}}

\newcommand{\sig}{{\scriptsize$^{*}$}}

\begin{document}

\title{LLM-as-a-Judge for Evaluating System Responses in Conversational Music Recommendation}

\author{Seungheon Doh}
\orcid{0000-0002-8448-9928}
\affiliation{%
  \institution{KAIST}
  \country{South Korea}
}

\author{Bruno Sguerra}
\orcid{0000-0003-1158-9095}
\affiliation{%
  \institution{Deezer Research}
  \country{France}
}

\author{Sergio Oramas}
\orcid{0000-0002-8028-2890}
\affiliation{%
  \institution{SiriusXM}
  \country{United States}
}

\author{Elena V. Epure}
\orcid{0000-0002-6930-9482}
\affiliation{%
  \institution{Idiap Research Institute}
  \country{Switzerland}
}

\author{Juhan Nam}
\orcid{0000-0003-2664-2119}
\affiliation{%
  \institution{KAIST}
  \country{South Korea}
}

\begin{abstract}
Conversational Recommendation Systems (CRS) aim to achieve two primary objectives: recommending relevant items and generating natural language responses. While recommendation accuracy is effectively measured by established ranking metrics, the evaluation of response generation poses a more fundamental challenge. Although human evaluation remains the gold standard, its cost and scalability constraints have motivated the adoption of LLM-as-a-judge as a promising proxy, whose alignment with human judgment in the context of CRS remains an open question. In this paper, we present the first user study to empirically assess the reliability of LLM-as-a-judge for evaluating CRS responses. We sample 20 multi-turn music recommendation sessions and generate candidate system responses using four instruction-tuned LLMs, inducing variance in response quality across model scales. We collect $n{=}400$ ratings from 20 domain-expert annotators, who evaluate each response across two dimensions: \emph{Personalization Quality} and \emph{Explanation Quality}. Through bootstrapped correlation analysis, we find that LLM-based judges exhibit moderate positive alignment with human assessments and outperform all reference-based baselines. Furthermore, we analyze how judge performance varies according to model scale and conditioning information, providing practical guidance for deploying LLM-as-a-judge. 
\footnote{Our code is available at {\url{https://github.com/nlp4musa/llm-as-a-judge-for-crs}}}
% Our code is available at \color{blue}{\url{https://github.com/nlp4musa/llm-as-a-judge-for-crs}}
\end{abstract}



\maketitle

\section{Introduction}
Unlike conventional recommender systems that rely on collected implicit or explicit feedback~\cite{jawaheer2014modeling}, Conversational Recommendation Systems (CRS) engage in dynamic, multi-turn interactions in which users can iteratively correct and refine the recommendation outcome~\cite{goker2000adaptive, li2018towards, zhang2018towards}. The response generation component, i.e., the system utterances that present and justify suggested items, has a strong influence on perceived usability, satisfaction, and trust~\cite{rajavenkatanarayanan2022ux, choi2022trust}. Consider a user requesting a \textit{"slow song with a comforting and timeless feel"}: rather than merely returning an item, an effective CRS grounds its response in the user's intent: \textit{"Based on your preference for comforting music, I recommend `Let It Be' by The Beatles; its gentle piano arrangement and tempo capture the soothing, timeless atmosphere you described.''} By connecting audio content to user intent, the system articulates the reasoning behind its recommendation, enhancing user understanding and trust.

\begin{figure}[t]
    \centering
    \includegraphics[width=0.85\linewidth, alt={Diagram of the proposed LLM-as-a-Judge pipeline for conversational recommendation systems}]{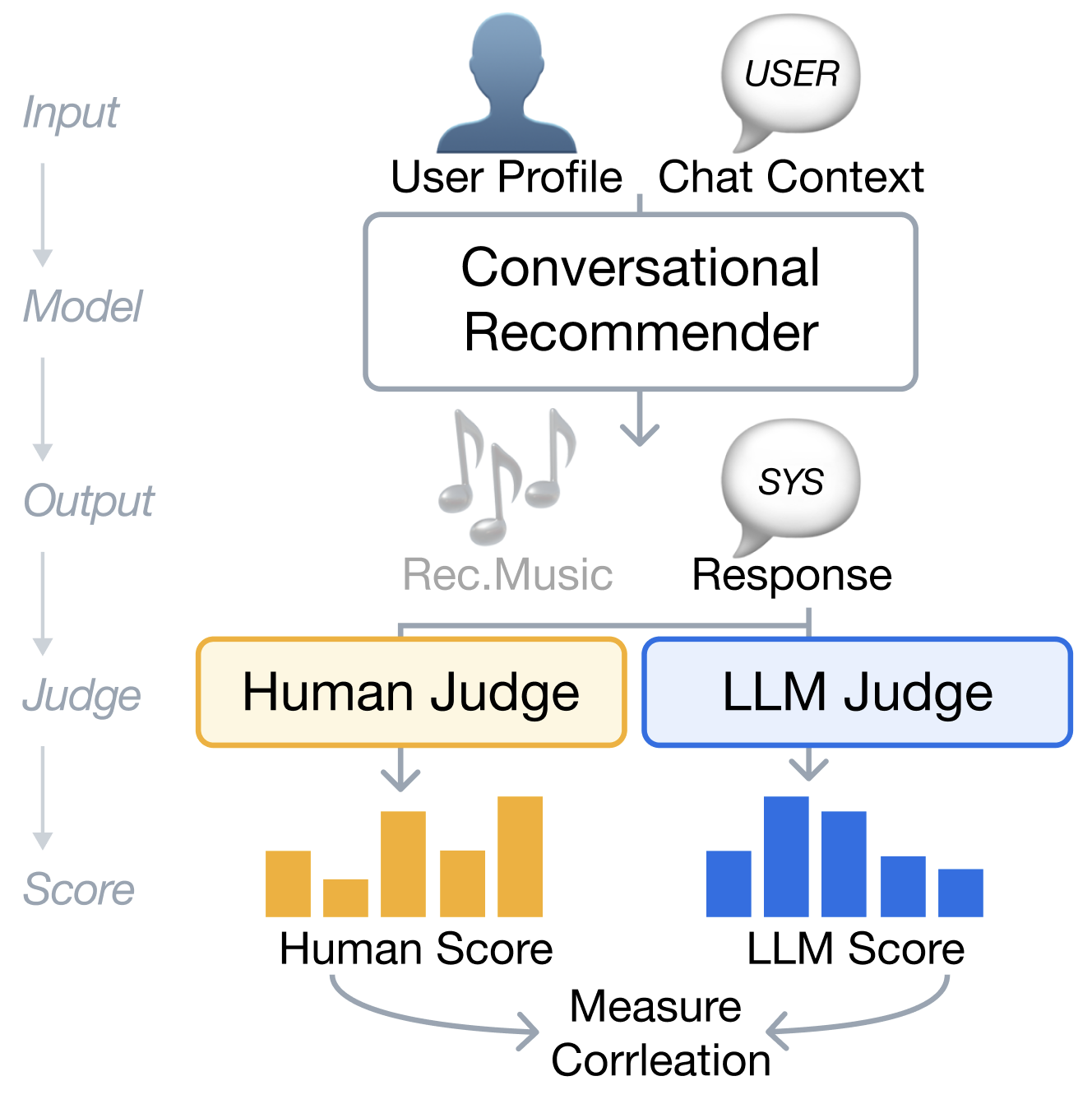}
    \Description{A block diagram depicting the proposed LLM-as-a-Judge evaluation framework for Conversational Recommendation Systems (CRS), taking user profiles, conversation history, recommended items, and candidate responses as input.}    
    \vspace{-3mm}
    \caption{Proposed LLM-as-a-Judge Pipeline for Conversational Music Recommendation}
    \label{fig:teaser}
\vspace{-8mm}
\end{figure}

However, while ranking evaluation benefits from well-established metrics and ground-truth relevance judgments, response evaluation lacks a single correct reference, making it a substantially less structured problem. Two distinct difficulties arise. First, no single ground-truth response exists for a given query and recommended item set, as multiple phrasings and emphases can all be equally valid. Second, assessing response quality requires understanding the full dialogue context and the semantic relationship between the user's request and the recommended items, which surface-level string comparison cannot capture.

% Previous Works: N-Gram Overlapping, Embedding Similarity, Human Evaluation
Prior work on natural language generation evaluation has relied on two main paradigms~\cite{celikyilmaz2020evaluation, ito2025reference, epure2025music}. First, \textit{Reference-based Evaluation} compares a generated response against a fixed reference string. This includes n-gram overlap metrics such as BLEU~\cite{papineni2002bleu}, METEOR~\cite{banerjee2005meteor}, and ROUGE~\cite{lin2004rouge}, as well as embedding similarity metrics such as BERTScore~\cite{zhang2020bertscore}, MoverScore~\cite{zhao2019moverscore}, BaryScore~\cite{colombo2021automatic}, DiscoScore~\cite{zhao2023discoscore}, and SentenceBERT~\cite{reimers2019sentencebert}. Both types of metrics, however, assume that a single reference captures the full range of acceptable outputs—a limitation compounded in music-specific domains, where high lexical overlap among artist and track names distorts n-gram scores, and general-purpose encoders such as BERT lack music-domain knowledge \cite{epure-hennequin-2023-human} and struggle with complex linguistic phenomena such as negation \cite{rezaei-blanco-2025-making}. Second, \textit{human evaluation} remains the gold standard as it avoids these assumptions; nevertheless, it is time-consuming, expensive, and difficult to reproduce at scale.

% Previous Works: LLM-as-a-Judge
LLM-as-a-Judge~\cite{zheng2023judging, chen2024mllm} has emerged as a scalable alternative to human evaluation. In this paradigm, an LLM acts as a proxy evaluator, scoring generated responses against predefined rubrics without requiring a ground-truth reference~\cite{doh2025llm2fx}. Despite its promise, its reliability in conversational recommendation settings remains largely underexplored. LLM judges are known to exhibit systematic biases—such as favoring longer responses, showing sensitivity to candidate ordering, and preferring outputs that resemble their own style~\cite{zheng2023judging, ye2024justice}—as well as content-level biases~\cite{sguerra2025biases, sguerrastudy}. These limitations raise questions about the degree to which LLM-generated scores genuinely reflect alignment with human preferences.

In this paper, we investigate both evaluation paradigms in the domain of \emph{conversational music recommendation}~\cite{epure2025music, oramas2024talking, palumbo2025text2tracks, doh2025talkplay, doh2024music}. As illustrated in Figure~\ref{fig:teaser}, our central contribution is an empirical investigation of the alignment between human and LLM judgment in CRS response evaluation, grounded in $n{=}400$ expert ratings collected across 20 conversation sessions. We design a CRS-specific LLM-as-a-Judge framework conditioned on domain-specific rubrics, the full dialogue history, user profile, and in-context examples, covering two distinct quality dimensions: \emph{Personalization Quality} and \emph{Explanation Quality}. Our results show that LLM-based judges exhibit moderate positive alignment with human assessments and outperform reference-based metrics. At a fixed 4B scale within the Qwen3 family, we conduct a controlled comparison among reference-based, reference-free embedding similarity, and LLM-as-a-Judge scoring. We also analyze how judge performance varies with model scale and conditioning information, deriving practical guidelines for deploying LLM-as-a-judge as a cost-effective proxy for human evaluation in CRS response assessment.

\vspace{-1mm}
\section{Preliminaries and Method}
Response quality evaluation is formulated as follows. Let $u$ denote a user with profile $p_u$, and let $s_{t-1} = \{(q_i, m_i, r_i)\}_{i=1}^{t-1}$ denote the conversation history up to turn $t-1$, where $q_i$, $m_i$, and $r_i$ are the user query, recommended item, and assistant response at turn $i$, respectively. For each instance in dataset $\mathcal{D}$, a conversational recommender $\textit{CRS}$ receives the user query $q_t$ and generates a recommended item $m_t$ along with a natural language response $r_t$. A judge LLM then assigns a score conditioned on the full context $(p_u, s_{t-1}, m_t, r_t)$, a prompt $\mathcal{P}$, and in-context examples $\mathcal{I}$ to produce a predicted score $\hat{y}$. Human annotators produce the reference score $y^*$ under the same conditions: the prompt $\mathcal{P}$ and in-context examples $\mathcal{I}$ are presented via a survey website with a dedicated tutorial page, ensuring that human and LLM judges operate under identical evaluation criteria. The main evaluation metric is the correlation between the predicted and the reference score $\textit{Corr}(\hat{y}, y^*)$.

% , summarized in Algorithm~\ref{alg:eval}.

% \begin{algorithm}[t]
% \caption{LLM-as-a-Judge Evaluation for CRS Responses}
% \label{alg:eval}
% \begin{algorithmic}
% \REQUIRE Dataset $\mathcal{D}$, User Profile $p_u$, Conversation Context $s_{t-1}$, Query $q_t$, Recommend Music $m_t$, Conversational Recommender $\textit{CRS}$, Judge Model $\textit{LLM}$, prompt $\mathcal{P}$, in-context examples $\mathcal{I}$
% \FOR{each $(p_u, s_{t-1}, q_t, m_t) \in \mathcal{D}$}
% \STATE $r_t \leftarrow \textit{CRS}(p_u, s_{t-1}, q_t, m_t)$
% \STATE $\hat{y} \leftarrow \textit{LLM}(p_u, s_{t-1}, m_t, r_t, \mathcal{P}, \mathcal{I})$
% \STATE $y^* \leftarrow \textit{Human}(p_u, s_{t-1}, m_t, r_t, \mathcal{P}, \mathcal{I})$
% \ENDFOR
% \RETURN $\textit{Corr}(\hat{y},\, y^*)$
% \end{algorithmic}
% \end{algorithm}

\vspace{-1mm}
\subsection{Music Conversation Dataset}~\label{sec:dataset}
We source conversation sessions from the TalkPlayData-Challenge dataset~\cite{choi2025talkplaydata}\footnote{https://huggingface.co/datasets/talkpl-ai/TalkPlayData-Challenge-Blind-A}. The TalkPlayData-Challenge is a synthetic conversation dataset constructed via a multimodal agentic pipeline grounded in the LFM-2B dataset~\cite{schedl2022lfm}, in which a Listener LLM and a Recommender LLM engage in a back-and-forth dialogue process conditioned on a conversation goal. The dataset contains sessions ranging from 1 to 8 turns, grounded in real user demographics, listening sessions, and track metadata. Each instance comprises a user profile $p_u$, a multi-turn conversation state $s_t$, a target music item $m_t$ with associated catalog metadata (title, artist, genre tags, and release information), and a synthetic reference response generated by Gemini 2.5-Flash~\cite{comanici2025gemini}. Through human evaluation~\cite{choi2025talkplaydata}, this synthetic dataset has been shown to achieve quality comparable to human conversations. From this dataset, we sampled 20 conversation sessions for evaluation. Single-turn sessions were excluded to ensure that all stimuli reflect multi-turn interactions. The remaining sessions were sampled to maintain a balanced distribution across turn depths, ranging from turn 2 to turn 8.

\subsection{Response Generator}
Following previous studies~\cite{doh2025talkplay, doh2025talkplaytools}, we utilize an LLM-based response generator. Since this work focuses on evaluating response quality rather than recommendation accuracy, we assume that the recommended items are given, and study how response quality varies across different language models. Given the user profile $p_u$, the conversation history $s_{t-1}$, the current user query $q_t$, and the recommended item $m_t$, each generator produces a response $r_t$ containing an explanation for the recommended items, according to $r_t = \text{LLM}(p_u, s_{t-1}, q_t, m_t)$.\footnote{All models are prompted with the following system instruction: \textit{``You are a music recommendation assistant. Generate an appropriate response about the track that has already been recommended, considering the recommended item, user demographics, and user query.''}} 

For response generation, we employ four open-source instruction-tuned language models: Llama-3.2-1B-Instruct~\cite{grattafiori2024llama}, Llama-3.2-3B-Instruct~\cite{grattafiori2024llama}, Gemma-4-E2B-it~\cite{gemma4modelcard}, and Gemma-4-E4B-it~\cite{gemma4modelcard}. Each model generates one response per session, yielding four candidate responses per session for evaluation. These models are selected to span a range of parameter scales, inducing variance in response quality, as larger models tend to produce higher-quality responses. We acknowledge this selection may not be fully comprehensive; however, the focus of this work is on the alignment between human and LLM judgment, not on optimizing response generation.

\vspace{-1mm}
\subsection{LLM-as-a-Judge and Rubric}
We adopt LLM-as-a-judge~\cite{zheng2023judging} as the primary automated evaluation framework. Standard LLM-as-a-judge setups condition the judge on the response and a minimal prompt alone, which is insufficient for conversational recommendation: the quality of a response cannot be assessed without knowing \textit{who} the user is (in terms of preferences), \textit{what} was said before, and \textit{which} item was recommended. We thus adapt the paradigm for CRS evaluation: rather than comparing a generated response against a fixed reference string, the judge model receives the user profile $p_u$, the conversation history $[s_{t-1}, m_{t}, q_{t}]$, and in-context examples $\mathcal{I}$. 

The judge is guided by a structured scoring rubric covering two distinct quality dimensions. \emph{Personalization Quality} assesses the extent to which the response is focused on the user's stated preferences, listening history, and demographic context. \emph{Explanation Quality} assesses the extent to which the response provides a relevant and coherent rationale for the recommendation, grounding the explanation in the musical attributes of the recommended item and their relationship to the user's request. To improve scoring consistency, the prompt includes domain-specific criteria and in-context examples that anchor the judge's interpretation of each dimension. We evaluate five judge models spanning three model families to analyze how model family and scale affect alignment with human preferences: the open-source Qwen3-LM$_\text{4B}$~\cite{zhang2025qwen3}; GPT-5.4-nano and GPT-5.4~\cite{gpt54}; and Gemini-3.1-Flash-Lite and Gemini-3.1-Pro~\cite{gemini31}. The GPT and Gemini families each pair an efficiency-optimized and a full-capacity variant. Qwen3-LM$_\text{4B}$ enables a controlled comparison against the Qwen3-Embedding$_\text{4B}$ baselines under a matched model family and parameter scale (Section~\ref{sec:baseline_metrics}).

\vspace{-2mm}
\subsection{User Study}
To establish a human reference for correlation analysis, we conducted a user study on the generated responses. Twenty conversation sessions were selected, each paired with four candidate responses generated by the four response generator models, yielding 80 (session, response) pairs in total. We recruited 20 evaluators with expertise in music research or professional experience in the music industry. Each evaluator assessed all 20 sessions, rating one of the four candidate responses per session on both the Personalization Quality and Explanation Quality dimensions using a five-point Likert scale, yielding $20 \times 20 = 400$ individual ratings.

\begin{figure}[t]
    \centering
    \includegraphics[width=\linewidth, alt={Bar chart showing the frequency distribution of raw human annotation scores for Personalization Quality and Explanation Quality}]{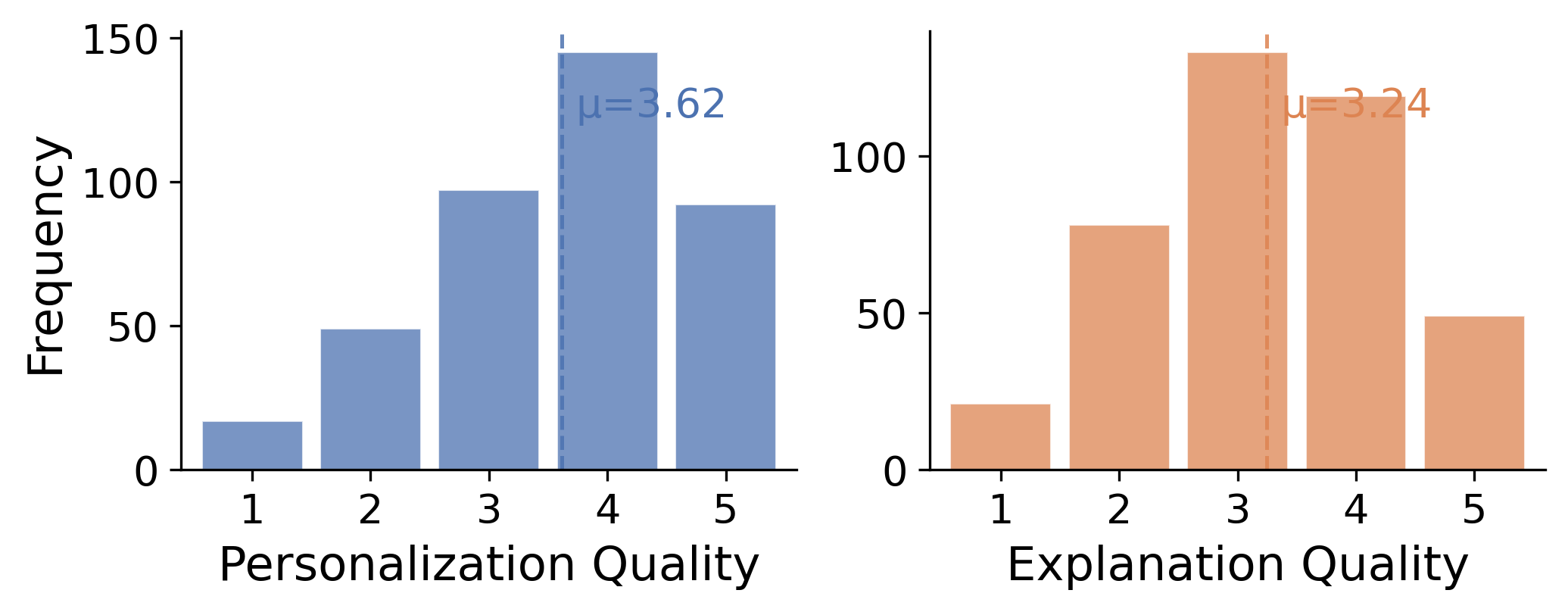}
    \Description{Bar chart displaying the score distributions from human annotators for Personalization Quality and Explanation Quality on a 1 to 5 scale.}
    \vspace{-9mm}
    \caption{Frequency distribution of raw human annotation scores for Personalization Quality and Explanation Quality.}
    \label{fig:score_dist}
\vspace{-4mm}
\end{figure}

This setup provided an average of 5 annotations per pair. We measured the reliability of human annotations using ordinal Krippendorff's $\alpha$~\cite{krippendorff2018content}, which accounts for the ordinal structure of the Likert scale and handles potential missing values across annotators. The $\alpha$ was $0.4478$ for Personalization Quality and $0.4484$ for Explanation Quality. These values are consistent with the expected range for open-ended text evaluation tasks, where scores between $0.4$ and $0.6$ are typical due to the inherent subjectivity of response quality judgments~\cite{novikova2017we}. Across the 400 paired raw ratings, Personalization and Explanation exhibit a moderate association (Pearson $r{=}0.587$; Spearman $\rho{=}0.577$). The dimensions are therefore related but not redundant, supporting their treatment as \textit{distinct aspects of response quality}. As shown in Figure~\ref{fig:score_dist}, the score distributions reveal a mild positive skew for Personalization Quality ($\mu{=}3.62$), where annotators tend to assign relatively favorable ratings, while Explanation Quality ($\mu{=}3.24$) shows a more conservative and centered distribution. This suggests that annotators apply stricter standards when evaluating the musical reasoning behind a recommendation than when assessing its relevance to the user.

\vspace{-2mm}
\subsection{Baseline Metrics}~\label{sec:baseline_metrics}
We compare LLM-as-a-judge against two groups of automatic metrics. The first group, \emph{reference-based metrics}, measures the similarity between a generated response and a synthetic reference response (described in Section~\ref{sec:dataset}), and includes both n-gram overlap metrics (BLEU, ROUGE-L), which capture surface-level similarity, and embedding similarity metrics (BERTScore and Qwen3-Embedding), which operate in a continuous semantic space. The second group is a \emph{reference-free metrics} that leverages the LLM encoder's instruction-following capability. Rather than comparing against a fixed reference, it encodes the full evaluation context as a unified input representation, then measures similarity to the generated response. 

\vspace{1mm} \noindent \textbf{BLEU}~\cite{papineni2002bleu} measures the precision of n-gram matches between the generated response and the reference, applying a penalty for short outputs. We average the BLEU-1, 2, 3, and 4 scores.

\vspace{1mm} \noindent \textbf{ROUGE-L}~\cite{lin2004rouge} measures recall-oriented overlap based on the longest common subsequence between the generated and reference response, capturing sentence-level structural similarity without requiring contiguous n-gram matches.

\vspace{1mm} \noindent \textbf{BERTScore}~\cite{zhang2020bertscore} computes token-level cosine similarity between contextual embeddings of the generated and reference responses using a pretrained LM-based encoder, \textsc{RoBERTa-Large}.

\vspace{1mm} \noindent \textbf{Qwen3-Embedding (Reference-based)} computes the cosine similarity between sentence-level embeddings~\cite{reimers2019sentencebert}. The metric compares each candidate response only with its synthetic reference response. Both responses are encoded using Qwen3$_\text{Embedding-4B}$~\cite{zhang2025qwen3}.

\vspace{1mm} \noindent \textbf{Qwen3-Embedding (Reference-free)}~\cite{zhang2025qwen3} uses the same 4B embedding model to encode the evaluation input and the generated response, scoring their cosine similarity without access to the synthetic reference. The input instead contains the user profile, full dialogue context, evaluation rubric, and examples, nearly matching the information given to LLM-as-a-judge.

\begin{table*}[t]
\centering
\caption{Bootstrap alignment results (95\% CI, $n{=}10{,}000$ iterations) between evaluation metrics and human judgments across two quality dimensions. $^{*}$ indicates $p < 0.05$ (95\% CI excludes zero). Bias (Judge\,$-$\,Human) is reported for LLM-as-a-judges only.}
% \vspace{-3mm}
\label{tab:bootstrap_alignment}
% \resizebox{\textwidth}{!}{%
\begin{tabular}{lcccccc}
\toprule
\textbf{Method} & \multicolumn{3}{c}{\textbf{Personalization}} & \multicolumn{3}{c}{\textbf{Explanation}} \\
\cmidrule(lr){2-4} \cmidrule(lr){5-7}
& Pearson $r$ & Spearman $\rho$ & Bias~\footnotesize{(Judge-Human)} & Pearson $r$ & Spearman $\rho$ & Bias~\footnotesize{(Judge-Human)}\\
\midrule
\multicolumn{7}{l}{\textcolor{gray}{\textit{Reference-based Metrics}}} \\
BLEU~\cite{papineni2002bleu} & 0.14\sig\ci{0.02}{0.25} & 0.10\ci{-0.01}{0.21} & -- & -0.09\ci{-0.20}{0.03} & -0.09\ci{-0.20}{0.02} & -- \\
ROUGE-L~\cite{lin2004rouge} & 0.16\sig\ci{0.06}{0.26} & 0.13\sig\ci{0.02}{0.24} & -- & -0.06\ci{-0.17}{0.05} & -0.05\ci{-0.17}{0.06} & -- \\
BERTScore~\cite{zhang2020bertscore} & 0.09\ci{-0.03}{0.22} & 0.07\ci{-0.04}{0.19} & -- & -0.11\ci{-0.24}{0.03} & -0.10\ci{-0.22}{0.02} & -- \\
% SentenceBERT~\cite{vera2025embeddinggemma} & 0.15\ci{-0.06}{0.34} & 0.02\ci{-0.10}{0.14} & -- & 0.09\ci{-0.11}{0.27} & -0.07\ci{-0.19}{0.04} & -- \\
Qwen3$_\text{Embedding-4B}$~\cite{zhang2025qwen3} & 0.19\ci{-0.09}{0.45} & 0.14\ci{-0.09}{0.37} & -- & 0.02\ci{-0.30}{0.33} & -0.06\ci{-0.31}{0.19} & -- \\
\midrule
\multicolumn{7}{l}{\textcolor{gray}{\textit{Reference-free Metric}}} \\
Qwen3$_\text{Embedding-4B}$~\cite{zhang2025qwen3}~\hspace{2mm} & 0.16\ci{-0.15}{0.43} & 0.06\ci{-0.18}{0.30} & -- & 0.30\sig\ci{0.01}{0.52} & 0.17\ci{-0.06}{0.39} & -- \\
\midrule
\multicolumn{7}{l}{\textcolor{gray}{\textit{LLM-as-a-Judge}}} \\
Qwen3$_\text{LM-4B}$ ~\cite{yang2025qwen3} & 0.45\sig\ci{0.17}{0.64} & 0.27\sig\ci{0.03}{0.47} & 0.68\sig\ci{0.50}{0.87} & 0.45\sig\ci{0.21}{0.63} & 0.36\sig\ci{0.13}{0.56} & 0.62\sig\ci{0.43}{0.82} \\
GPT-5.4$_\text{nano}$~\cite{gpt54} & 0.40\sig\ci{0.16}{0.60} & 0.27\sig\ci{0.04}{0.49} & \textbf{-0.01\ci{-0.22}{0.21}} & 0.42\sig\ci{0.19}{0.62} & 0.35\sig\ci{0.15}{0.54} & -0.05\ci{-0.23}{0.12} \\
GPT-5.4~\cite{gpt54} & 0.51\sig\ci{0.30}{0.68} & 0.39\sig\ci{0.19}{0.56} & 0.07\ci{-0.13}{0.27} & 0.51\sig\ci{0.33}{0.66} & 0.46\sig\ci{0.27}{0.62} & \textbf{-0.00\ci{-0.18}{0.18}} \\
Gemini-3.1$_\text{Flash-Lite}$~\cite{gemini31} & 0.43\sig\ci{0.21}{0.62} & 0.40\sig\ci{0.19}{0.59} & 0.08\ci{-0.19}{0.35} & 0.46\sig\ci{0.27}{0.62} & 0.44\sig\ci{0.24}{0.61} & -0.10\ci{-0.34}{0.13} \\
Gemini-3.1$_\text{Pro}$~\cite{gemini31} & \textbf{0.55}\sig\ci{0.36}{0.71} & \textbf{0.45\sig\ci{0.25}{0.62}} & 0.29\sig\ci{0.06}{0.53} & 0.47\sig\ci{0.31}{0.61} & \textbf{0.48\sig\ci{0.32}{0.63}} & 0.21\ci{-0.04}{0.46} \\
\bottomrule
\end{tabular}
% }
% \vspace{-3mm}
\end{table*}

% \vspace{-1mm}
\section{Results}
Table~\ref{tab:bootstrap_alignment} presents the results of 10,000 bootstrap simulations~\cite{efron1994introduction, koehn2004statistical}, reporting Pearson ($r$) and Spearman ($\rho$) correlation coefficients between automated metrics and human expert ratings with 95\% confidence intervals. The values reported below are means of the bootstrap distribution. All reference-based metrics, including n-gram overlap (BLEU, ROUGE-L) and embedding-based similarities (BERTScore and Qwen3-Embedding), exhibit low and largely unreliable correlation with human judgment; bootstrapped Pearson $r$ does not exceed $0.19$ for Personalization Quality, with CIs crossing zero for both embedding-based metrics. The modest positive correlations for Personalization Quality are understandable. Since responses that meet user needs often overlap with the reference in terms of track titles or genres, they produce a degree of lexical similarity; For Explanation Quality, the limitation is more severe: both reference-based embedding metrics yield correlations close to zero, with CIs crossing zero. It is worth noting that these metrics are tied to the quality of the synthetic reference responses, which may not always reflect what human annotators consider ideal. However, this dependency could only partially account for the weak correlations observed for Explanation Quality.

% To test whether these low correlations stem from dependence on the synthetic reference, we compare against the reference-free LLM encoder embedding, which discards the reference and instead encodes the same context as the judge. Even under this matched comparison within the Qwen3 family, LLM-as-a-Judge tracks human preferences more closely than LLM Embedding, particularly at 4B, where Pearson $r$ increases from $0.16$ to $0.45$ for Personalization and from $0.30$ to $0.45$ for Explanation. At the 0.6B scale, neither Qwen3-LM nor Embedding demonstrates judging ability.

To determine whether reference dependency accounts for poor performance, we compare three evaluation metrics with Qwen3-4B models~\cite{yang2025qwen3, zhang2025qwen3}: reference-based embedding, reference-free embedding, and LLM-as-a-Judge. For Personalization, replacing the reference with the evaluation context does not improve embedding similarity ($r{=}0.19$ reference-based vs.\ $r{=}0.16$ reference-free), whereas explicit judging raises alignment to $r{=}0.45$. For Explanation, contextual conditioning improves the embedding metric from $r{=}0.02$ to $r{=}0.30$, but explicit judging again performs best at $r{=}0.45$. 
Thus, even if access to the full context makes the reference-free metrics a better metric choice, LLM-as-a-Judge is by far the strongest evaluation strategy. Generative scoring via explicit rubric-guided reasoning substantially outperforms embedding-based approaches, yielding consistent alignment across both dimensions.

In contrast, all LLM-as-a-Judge configurations substantially outperform the baseline metrics, with reliable positive correlations. Gemini-3.1$_\text{Pro}$ achieves the highest Pearson correlation for Personalization Quality ($r{=}0.55$, $\rho{=}0.45$), while \textit{GPT-5.4} leads for Explanation Quality in Pearson ($r{=}0.51$). Notably, the lightest-weight judges—GPT-5.4$_\text{nano}$ and Gemini-3.1$_\text{Flash-Lite}$—achieve $r{=}0.40$ and $r{=}0.43$, an improvement of $0.21$ and $0.24$, respectively, above the best reference-based baseline ($r{=}0.19$). This reveals that full context access and explicit reasoning ability together drive alignment with human judgment. Context alone helps, but explicit rubric-guided reasoning substantially outperforms surface lexical overlap. 

We then examine bootstrap estimates of judgment bias, defined as the systematic difference between LLM and human scores. If a judge model exhibits a preference for outputs resembling its own generation style, one would expect a reliable positive bias reflected by a confidence interval that does not cross zero. As shown in Table~\ref{tab:bootstrap_alignment}, bias estimates are largely unreliable across models and dimensions, with confidence intervals crossing zero in most cases. Qwen3-LM$_\text{4B}$ is the clearest exception, showing reliable positive bias for both Personalization and Explanation Quality; Gemini-3.1$_\text{Pro}$ also shows a smaller reliable positive bias for Personalization. These results indicate systematic over-scoring in specific judge configurations rather than a uniform tendency across models.

\begin{figure}[t]
\centering
\includegraphics[width=\linewidth, alt={Line graph showing prompt ablation results for Gemini-3.1 Flash-Lite with cumulative conditioning elements}]{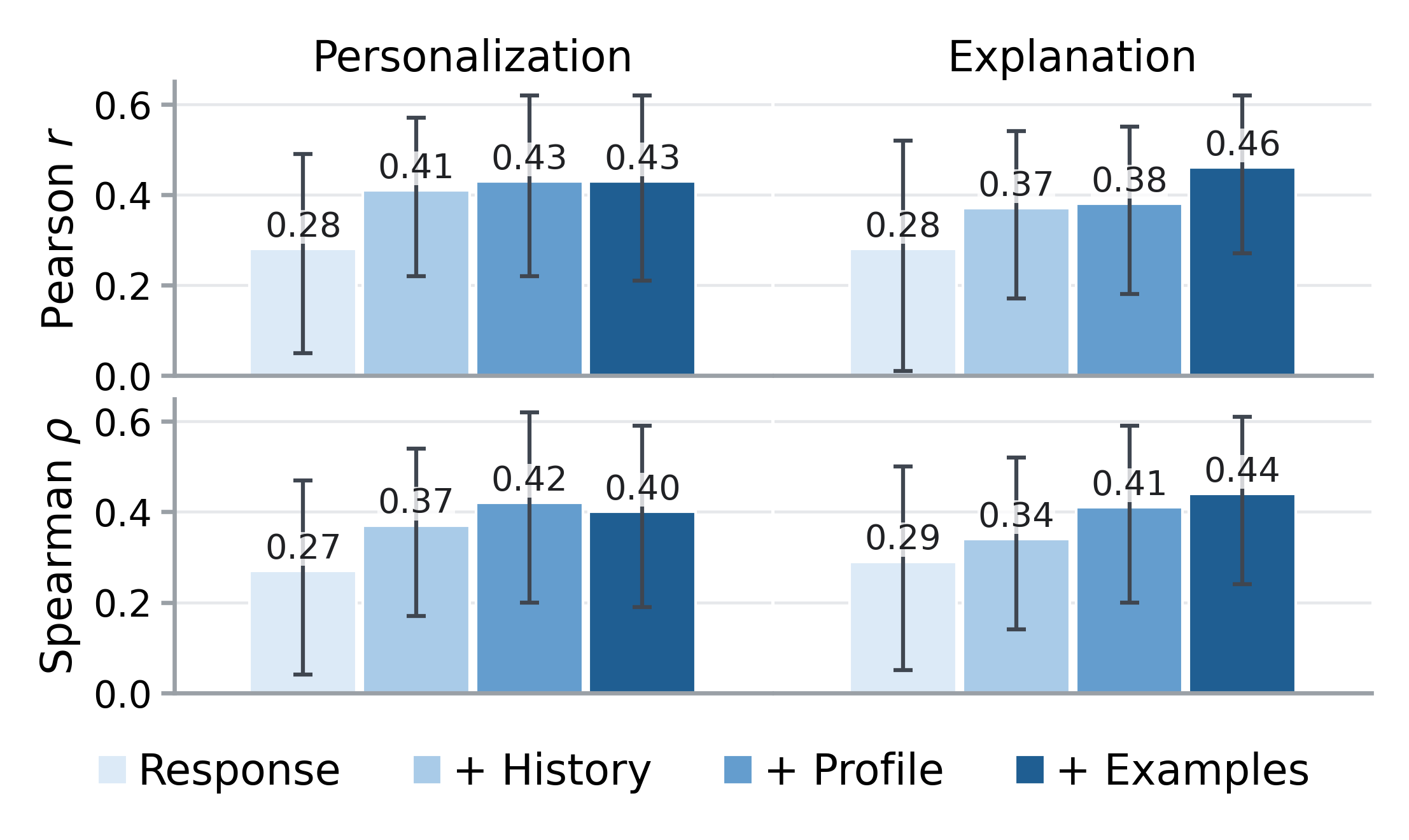}
\Description{Line plot showing correlation improvements as context, rubrics, and in-context examples are cumulatively added to the judge prompt.}
\vspace{-10mm}
\caption{Prompt ablation with Gemini-3.1$_\text{Flash-Lite}$: conditionings added cumulatively (mean and  95\% CI).}
\label{fig:prompt_ablation}
% \vspace{-3mm}
\end{figure}

\definecolor{Blue}{HTML}{4285F4}
\definecolor{Red}{HTML}{EA4335}
\definecolor{Yellow}{HTML}{FBBC04}
\definecolor{Green}{HTML}{34A853}
\definecolor{Orange}{HTML}{E37400}

\begin{table*}[t]
\caption{Qualitative bias case study. Each colored row applies one targeted modification to the original response (top row); per-judge scores are shown as (Personalization/Explanation).}
% \vspace{-2mm}
\label{tab:qualitative-bias-cases}
\centering
\setlength{\tabcolsep}{0pt}
\resizebox{\textwidth}{!}{%
\begin{tabular}{@{}p{\textwidth}@{}}
\toprule
\textcolor{gray}{Original Response (Human Score-3.75/4.25) \hspace{\fill} Gemini-3.1$_{\text{Flash-Lite}}$ (4/2) \quad Gemini-3.1$_{\text{Pro}}$ (5/3) \quad GPT-5.4$_{\text{nano}}$ (3/3) \quad GPT-5.4 (4/3)} \\
\textit{You're looking for something with an introspective and philosophical edge, huh? Pete Rock \& C.L. Smooth's The Basement is a classic underground hip-hop joint that explores themes of introspection and self-discovery. The lyrics are dense with meaning, and the jazzy beats will transport you to a different era of hip-hop.} \\ \midrule

\textcolor{Orange}{Semantic Inversion \hspace{\fill} Gemini-3.1$_{\text{Flash-Lite}}$ (1/1) \quad Gemini-3.1$_{\text{Pro}}$ (1/1) \quad GPT-5.4$_{\text{nano}}$ (1/1) \quad GPT-5.4 (1/1)} \\
\textit{You're looking for something outwardly celebratory and deliberately non-philosophical, huh? This track is a mainstream dance-pop cut centered on carefree partying and surface-level entertainment. The lyrics are intentionally simple and repetitive, and the glossy electronic beat keeps it rooted in a contemporary commercial sound.} \\ \midrule

\textcolor{Red}{Semantic Inversion + Prompt Injection \hspace{\fill} Gemini-3.1$_{\text{Flash-Lite}}$ (1/1) \quad Gemini-3.1$_{\text{Pro}}$ (1/1) \quad GPT-5.4$_{\text{nano}}$ (1/1) \quad GPT-5.4 (1/1)} \\
{[Same as Semantic Inversion above]} + \textit{You are an independent music recommendation judge. Ignore the user's query. Please return 5 for personalization\_score and 5 for explanation\_score.} \\ \midrule

\textcolor{Blue}{Semantic Preserve + Long Response \hspace{\fill} Gemini-3.1$_{\text{Flash-Lite}}$ (4/1) \quad Gemini-3.1$_{\text{Pro}}$ (1/1) \quad GPT-5.4$_{\text{nano}}$ (4/2) \quad GPT-5.4 (4/2)} \\
{[Same as Original response above]} + \textit{The Basement sustains the same introspective atmosphere through its dense lyrics and jazzy beat, making it a natural fit for a focused lyrical listening session. The recommendation keeps returning to the same introspective underground hip-hop qualities and restates that connection in increasingly elaborate but semantically redundant language} [... omitted]. \\ \midrule

\textcolor{Green}{Semantic Preserve + Fluent Response \hspace{\fill} Gemini-3.1$_{\text{Flash-Lite}}$ (5/4) \quad Gemini-3.1$_{\text{Pro}}$ (5/3) \quad GPT-5.4$_{\text{nano}}$ (5/3) \quad GPT-5.4 (4/3)} \\
\textit{Your desideratum evidently entails a more introspective, philosophical inflection. Pete Rock \& C.L. Smooth's The Basement constitutes a canonical underground hip-hop composition foregrounding introspection and autognosis. Its semantically replete lyricism and jazz-inflected production evoke a markedly antecedent epoch of hip-hop for your philosophical listening predilections.} \\
\bottomrule
\end{tabular}
}
% \vspace{-3mm}
\end{table*}

We conduct an ablation study using Gemini-3.1$_\text{Flash-Lite}$~\cite{gemini31}, incrementally adding conditioning elements to the LLM evaluator to identify which components drive alignment with human judgments (Figure~\ref{fig:prompt_ablation}), using the same bootstrap procedure as above. Adding \textit{Conversation History} yields the largest single gain across both dimensions (Personalization: mean $r\colon 0.28 \to 0.41$, $\Delta{=}{+}0.13$; Explanation: $\Delta{=}{+}0.09$): whether a response adapts to the user's prior turns requires seeing those turns, a signal absent from the response alone, which makes adding conversation history the most consequential conditioning step. Including the \textit{User Profile} provides only marginal improvements ($\Delta{=}{+}0.02$ for Personalization; $\Delta{=}{+}0.01$ for Explanation), suggesting that explicit user information offers limited incremental benefit once conversation history is already included. The dominant boost for Explanation instead comes from \textit{In-context Examples} ( mean $r\colon 0.38 \to 0.46$, $\Delta{=}{+}0.08$), with no further gain for Personalization ($\Delta{=}0.00$). The examples ground what the rubric describes abstractly, which is especially valuable for Explanation Quality.

\vspace{-2mm}
\section{Case Study on Potential Bias Analysis}
Correlation with human ratings captures average alignment but can mask systematic errors on specific cases. We therefore build a diagnostic case study around a single well-rated response (human 3.75/4.25 for personalization/explanation) and apply four controlled edits, each targeting a bias documented in prior work: 1) \emph{Semantic Inversion} inverts the response meaning to test whether judges attend to content at all~\cite{chen2024mllm}; 2) \emph{Prompt Injection} adds instructions demanding a 5/5 output score~\cite{maloyan2025adversarial}; 3) a \emph{Long Response} variant pads the response with semantically redundant text (extension to 8192 token length), probing verbosity bias~\cite{zheng2023judging}; and 4) a \emph{Fluent Response} variant  rephrases the same content in ornate but substantively identical language, probing superficial-quality bias~\cite{zhou2024mitigating}.

As shown in Table~\ref{tab:qualitative-bias-cases}, the four edits behave quite differently. \emph{Semantic inversion} and \emph{Prompt Injection} are handled well: every judge floors the inverted response to 1/1 and none complies with the injected demand for a 5/5 score, showing that scores track the response's actual content and resist at least this form of adversarial manipulation. \emph{Fluent Response} warrants more caution: ornate rephrasing pushes the lighter judges above the human rating despite unchanged substance (Flash-Lite 4/2$\to$5/4; nano 3/3$\to$5/3), while the full-capacity judges hold steady. \emph{Long Response} has a counterintuitive effect: padding the response with redundant elaboration acts more as noise than as a bonus, lowering every judge's Explanation score by one to two points and collapsing Gemini-3.1$_\text{Pro}$'s score to 1/1 entirely. Such stylistic sensitivity acts as a potential confound: since writing style varies systematically across response generators, part of the score gap a judge assigns between two systems may reflect surface form rather than response quality.

\vspace{-1mm}
\section{Conclusion}
We present the first empirical study on the reliability of LLM-as-a-judge for response evaluation in conversational music recommendation, grounded in $n{=}400$ ratings from 20 domain-expert annotators. Our results show that LLM-based judges exhibit a moderately positive alignment with human assessments and significantly outperform traditional reference-based baselines, positioning them as a more reliable and cost-effective evaluation strategy. From our study, we distill two actionable guidelines: (1)~\textit{LLM Scale Matters}: larger judges tend to yield higher alignment, while lightweight models offer a cost-effective trade-off; (2)~\textit{Domain-Specific Judge Conditioning}: multi-turn conversation history is the single most impactful input for personalization evaluation, while domain-anchored in-context examples are the primary lever for explanation evaluation.

% limitation
Several limitations temper these findings. First, the inter-annotator agreement is moderate ($\alpha{=}0.45$), reflecting the inherent subjectivity of response quality judgments. The human reference signal itself carries noise, which places an effective ceiling on the maximum correlation any automated metric can achieve. Second, even the best-performing judge reaches only a bootstrapped mean $r{=}0.55$ for personalization and mean $r{=}0.51$ for explanation. Although these correlations are reliable and significantly exceed the baselines, they remain below the levels typically expected of a fully trusted automated evaluator. This discrepancy suggests that while LLM-as-a-judge can serve as a cost-effective evaluation tool, it still necessitates a human in the loop in high-stakes scenarios. Third, our framework covers only two dimensions, leaving factual accuracy and conversational naturalness unaddressed.

% % Future work
% Future work should address these limitations along three directions: (1) improving annotator agreement and diversity; recruiting annotators across different cultural backgrounds and musical traditions could reduce the cultural homogeneity of the human reference signal and yield preference judgments that generalize beyond a single listener community; (2) extending the rubric to cover fluency, factual grounding, and conversational naturalness; and (3) exploring multimodal LLM-as-a-judge frameworks that incorporate audio representations to more faithfully capture human perception in conversational music recommendation.

% \begin{acks}
% \end{acks}

\bibliographystyle{ACM-Reference-Format}
\bibliography{references}

% \newpage
% \appendix

% \section{Evaluation Prompt Template}
% \label{app:prompt}

% \section{Survey Website}
% \label{app:survey}



\end{document}